\documentclass{article}
\usepackage{spconf,amsmath,graphicx,multirow}
\usepackage{hyperref}
\usepackage{xcolor}
\usepackage{amssymb}

\title{Learned Quality Enhancement via Multi-frame Priors for HEVC Compliant Low-Delay Applications}
\name{Ming Lu$^\dag$, Ming Cheng$^\ddag$, Yiling Xu$^\ddag$, Shiliang Pu$^\star$, Qiu Shen$^\dag$, and Zhan Ma$^\dag$\thanks{Emails: luming@smail.nju.edu.cn, \{ming\_cheng, yl.xu\}@sjtu.edu.cn, pushiliang@hikvision.com, \{shenqiu, mazhan\}@nju.edu.cn}}
\address{$^\dag$Nanjing University, $^\ddag$Shanghai Jiao Tong University, $^\star$Hikvision Research Institute}
\begin{document}
%\ninept
%
\maketitle
\begin{abstract}
{Networked video applications, e.g., video conferencing, often suffer from poor visual quality due to unexpected network fluctuation and limited bandwidth. In this paper, we have developed a Quality Enhancement Network (QENet) to reduce the video compression artifacts, leveraging the spatial and temporal priors generated by respective multi-scale convolutions spatially and warped temporal predictions in a recurrent fashion temporally. We have integrated this QENet as a standard-alone post-processing subsystem to the High-Efficiency Video Coding (HEVC) compliant decoder. Experimental results show that our QENet demonstrates the state-of-the-art performance against default in-loop filters in HEVC and other deep learning based methods with noticeable objective gains in Peak-Signal-to-Noise Ratio (PSNR) and subjective gains visually.}
\end{abstract}
\section{Introduction}
\label{sec:intro}
Networked video applications prevail in our daily life. High-efficiency video compression is demanded to ensure the smooth network delivery and guarantee the satisfactory Quality of Experience (QoE) for end users. However, lossy video compression is always accompanied by undesired artifacts, such as blocky, motion blurring and ringing. In-loop filters (e.g., deblocking, and/or Sample Adaptive Offset (SAO)) are incorporated in popular HEVC~\cite{HEVC} standard to reduce the compression artifacts for quality improvement.

On the other hand,  deep Convolutional Neural Networks (CNNs) have been applied to alleviate the quantization induced compression artifacts, and demonstrated noticeable performance enhancements.
For example, Park and Kim \cite{park2016cnn} used a CNN based approach to replace SAO in HEVC. Dong {\it et al.} \cite{dong2015compression} built an Artifacts Reduction CNN (ARCNN)~\cite{dong2014learning} to reduce artifacts introduced by JPEG compression. Dai \emph{et al.} \cite{dai2017convolutional} proposed a Variable-size Residual learning CNN (VRCNN) to further improve the quality of HEVC compressed videos. In the meantime, Nah \emph{et al.}~\cite{nah2017deep} and Tao \emph{et al.}~\cite{tao2018scale} adopted deep multi-scale CNNs that utilize the local and non-local information within the single frame for image deblurring. Zhang {\it et al.}~\cite{zhang2017beyond} designed Denoising CNNs (DnCNNs) with very deep architecture for image denoising.

Temporal correlations or priors can be used for video related tasks, e.g., motion representation, frame interpolation, and video denoising. For example, Long Short-Term Memory (LSTM) \cite{xingjian2015convolutional} based algorithms were proposed to handle temporal processing tasks by transmitting intermediate parameters iteratively.  Dosovitskiy \emph{et al.} \cite{dosovitskiy2015flownet} and Ilg \emph{et al.} \cite{ilg2017flownet} proposed an optical flow network to predict the motion vector between consecutive frames via Deep Neural Network (DNN) rather than traditional block based motion search. Niklaus \emph{et al.} \cite{niklaus2017video} utilized learned adaptive convolutions to capture the motion accurately for better frame warping. With these flow based motion representations, temporal priors can be captured from previously reconstructed multiple frames to further the artifacts reduction (e.g., motion deblurring) for a single-frame quality enhancement approaches~\cite{dong2015compression}. Lu {\it et al.}~\cite{lu2018deep} introduced a Deep Kalman Filtering Network (DKFN) for video compression artifact reduction.

\begin{figure}[t]
\centering
\includegraphics[scale=0.34]{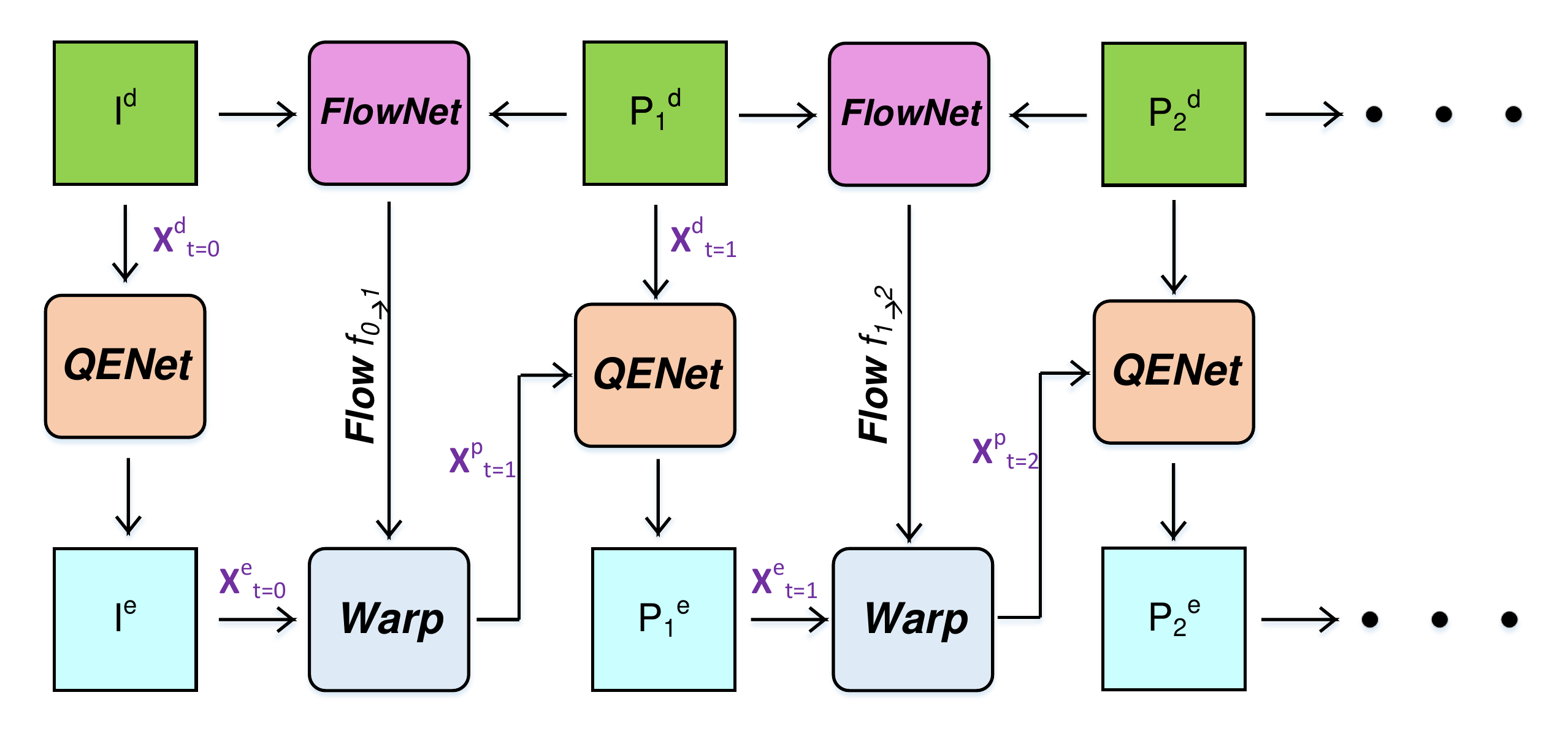}
\caption{Illustration of multi-frame QENet for a low-delay video application with temporal priors  captured recurrently.}
\label{sfig:framework}
\end{figure}

\begin{figure*}[t]
\centering
\includegraphics[scale=0.5]{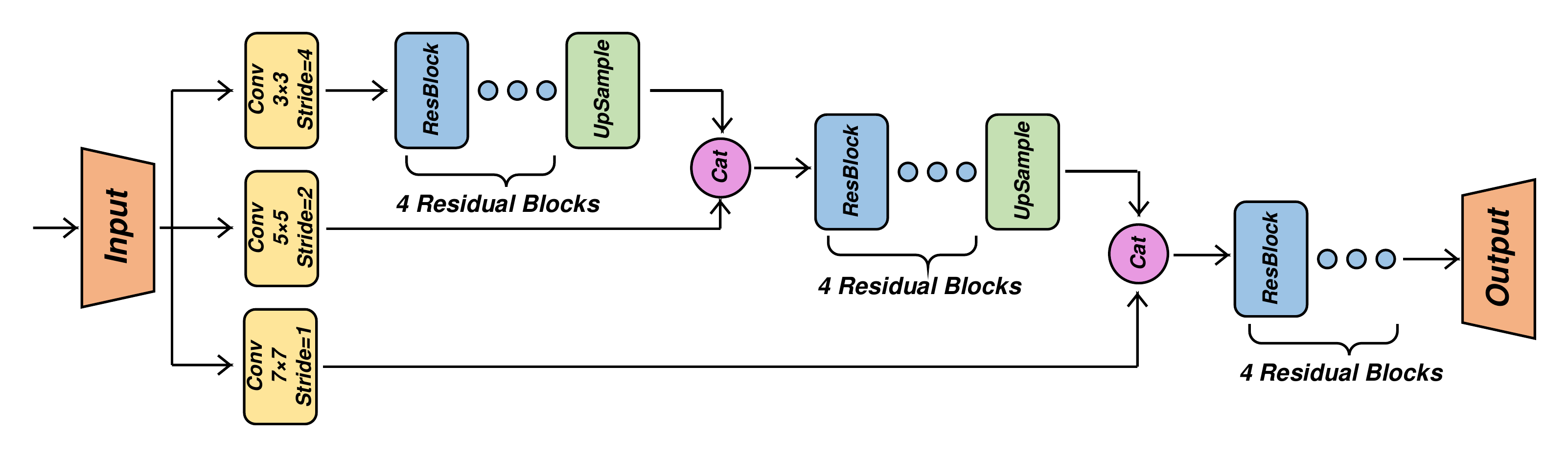}
\caption{Illustration of scale-wise convolutions to capture multi-scale spatial priors within current frame.}
\label{sfig:enhancement_network}
\end{figure*}

In this work, we first design an effective quality enhancement network with scale-wise convolutions to capture the multi-scale spatial priors within current frame for single-frame compression artifacts reduction. Such multi-scale convolutions coincides with the variable-size Coding Unit (CU) idea utilized in HEVC to well exploit the regional content characteristics (i.e., rich texture area with small-size convolution and CU, and stationary background with large-size convolution and CU).  In addition, we also explore the temporal relationships among consecutive frames to further improve the quality. Such temporal priors are generated in a recurrent way from previously reconstructed frames via optical flow estimation. We have trained our QENet in an end-to-end fashion for HEVC compliant compressed video enhancement as a standard-alone post-processing module. Experimental results show that our multi-frame QENet demonstrates the state-of-the-art performance against default filters in HEVC with about 1.8 dB PSNR gain, ARCNN~\cite{dong2015compression} with about 1 dB PSNR gain, DnCNN~\cite{zhang2017beyond} with about 0.4 dB PSNR gain and DKFN~\cite{lu2018deep} with about 0.24 dB PSNR gain, respectively, for a low-delay application with intermediate bit rate (e.g, compressed using quantization parameter QP 32). Similar gains are kept at QP 37 (i.e., low bit rate scenario).

\section{Learned Quality Enhancements Networks for Compressed Video}
\label{sec:format}
Figure~\ref{sfig:framework} presents our proposed multi-frame QENet for video quality enhancement.
We use HEVC compliant IPPP structure for a low-delay scenario representation. Our QENet works recurrently from the first I frame ${\bf X}^d_{t=0}$ to the upcoming P frames ${\bf X}^d_t, t = 1, 2, 3, \ldots$ consecutively, to produce frames ${\bf X}^e_t$ with enhanced quality.

\subsection{Single-Frame (SF) QENet}
\label{ssec:SF_QENet}

Since there is no {\it temporal priors} for I frame, our Single-Frame (SF) QENet only enhances it using multi-scale spatial priors, as shown in Fig.~\ref{sfig:enhancement_network} with only current frame as input, i.e.,
\begin{align}
{\bf X}^e_{t} = {\bf QENet}\left({\bf X}^d_t\right), \mbox{~~~}t = 0. \label{eq:sf_qenet}
\end{align}
We utilize scale-wise convolutions to capture the multi-scale priors spatially in current frame. %It is simply referred to as {\it SF QEnet}.
Different from \cite{nah2017deep} and \cite{tao2018scale}, the input frame of our SF-QENet is resized into three scales respectively using convolutions by setting stride at 1, 2 and 4 instead of applying typical down scaling filters (e.g., bilinear, bicubic). With such implementation, local and non-local spatial information of current frame can be extracted and fused effectively. Three different convolutional kernel sizes are utilized accordingly, e.g., $3\times3$ convolution for 1/16 down scaled size of the original content, $5\times5$ for 1/4 of the original size and $7\times7$ for the original scale. This helps to extract features from different scales. Four Residual Blocks (ResBlock) \cite{kim2016accurate} are applied at each scale for acquisition of high-dimensional features with the deconvolution upsampling layer connected afterwards. The number of channels is basically 64 of each convolutional layer. Under such connection architecture, spatial information of each scale can be fused together closely for final quality restoration and enhancement.

\subsection{Multi-Frame (MF) QENet} \label{ssec:MF_QENet}

For upcoming decoded P frames, optical flow $f_{(t-1)\rightarrow t}$ between consecutive decoded frames, i.e., ${\bf X}^d_{t-1}$ and ${\bf X}^d_{t},  t > 1$ , is generated via dedicated Flow estimation Network (FlowNet) $\mathbb F$~\cite{ilg2017flownet}. Flows will then be used to warp previously enhanced frames (e.g., ${\bf X}^e_{t-1}$) for intermediate frame prediction ${\bf X}^p_{t}$. Together with decoded ${\bf X}^d_t$, we could finally restore the ${\bf X}^e_t$, i.e.,
\begin{align}
f_{(t-1)\rightarrow t} &= {\mathbb F}\left({\bf X}^d_{t-1}, {\bf X}^d_t\right), \\
{\bf X}^p_{t} &= {\bf WARP}\left(f_{(t-1)\rightarrow t}, {\bf X}^e_{t-1}\right), \label{eq:warp_frame}\\
{\bf X}^e_{t} &= {\bf QENet}\left({\bf CAT}\left({\bf X}^p_{t}, {\bf X}^d_t\right)\right), \mbox{~~~}t > 0. \label{eq:mf_qenet}
\end{align}

\begin{table*}[t]
\centering
%\footnotesize
\caption{Averaged (PSNR (dB), SSIM) using Vimeo Test Sequences}
\label{tab:results}
\begin{tabular}{|c|c|c|c|c|c|c|c|}
\hline
QP & HEVC & HEVC-LF & ARCNN & DnCNN & DKFN & SF-QENet & MF-QENet \\
\hline
32 & (33.87, 0.946) & (34.18, 0.950) & (35.08, 0.957) & (35.58, 0.961) & (35.81, 0.962) & (35.76, 0.962) & \textbf{(36.01, 0.964)} \\
\hline
37 & (31.64, 0.917) & (31.98, 0.923) & (32.68, 0.933) & (33.01, 0.936) & (33.23, 0.939) & (33.28, 0.940) & \textbf{(33.54, 0.944)} \\
\hline
\end{tabular}
\end{table*}

%\subsection{Optical flow network for motion estimation}
%\label{ssec:flow}
We modify the FlowNetS proposed in \cite{dosovitskiy2015flownet} as our optical flow network by replacing the bilinear upsampling with the pixel shuffle \cite{shi2016real} upsampling.
The U-net \cite{ronneberger2015u} architecture is adopted as the motion estimation network with six scales which is very effective for accurate motion capture. Batch normalization layers are necessary after each convolutional layer for such estimation network which may not be suitable for restoration networks however. Deconvolution layers are adopted to upsample the estimated flow which is fused together with feature maps of the input in the same scale to reconstruct the larger scale as in \cite{dosovitskiy2015flownet}. The final flow records the motion vectors of the two consecutive frames for subsequent warping with ${\bf X}^e_{t-1}$. Note that the FlowNetS may not be the best optical flow network but it is easier for training with less computational complexity.

In contrast to SF-QENet in Eq.~\eqref{eq:sf_qenet} with only current frame as input, our MF-QENet shown in Eq.~\eqref{eq:mf_qenet} takes inputs from both ${\bf X}^p_{t}$ and ${\bf X}^d_t$ that are concatenated with temporal priors embedded recurrently. Note that QENet used in either \eqref{eq:sf_qenet} and \eqref{eq:mf_qenet} also captures the multi-scale spatial priors for quality enhancements.

\subsection{Loss Function for End-to-End Learning}
\label{ssec:loss}
We use two loss functions to train our model. The $\mathcal{L}_2$  loss $L _{e} $ is applied in QENet as illustrated in Fig.~\ref{sfig:framework} and is back propagated through both QENet and FlowNet~\cite{ilg2017flownet}:
\begin{equation}
L_{e} = \frac{1}{n} \sum_{t=1}^n \left| {\bf X}^{e}_t - {\bf X}^{\rm org}_t \right|^2,
\end{equation} with ${\bf X}^e_t$ and ${\bf X}^{\rm org}_t$ indicating the enhanced frame and original frame at $t$.
To calculate the optical flow more precisely, we add another $\mathcal{L}_1$ norm $L_{w} $ between the estimated frame ${\bf X}^p_t$ warped from the previous restored frame via Eq.\eqref{eq:warp_frame} and its corresponding decoded frame ${\bf X}^d_t$:
\begin{equation}
L_{w} = \sum_{t=1}^n \left| {\bf X}^{p}_t - {\bf X}^{d}_t \right|
\end{equation}
The total loss used for end-to-end training is $ L = L_{e} + L_{w}  $.

\section{Experimental Studies}
\label{sec:experiments}
Our experiments were performed on a desktop with an i7-7700K CPU and a NVIDIA Quadro P5000 GPU. PyTorch \cite{ketkar2017introduction} platform was chosen to implement the proposed model.

%\subsubsection{Experimental settings}
%\label{sssec:settings}
%As listed in Table~\ref{tab:ffmpeg}, we generate compressed frames using ffmpeg (https://www.ffmpeg.org/) for training and evaluation. The parameters not listed in the table are kept at the default values.

\begin{figure*}[t]
\centering
\includegraphics[scale=0.65]{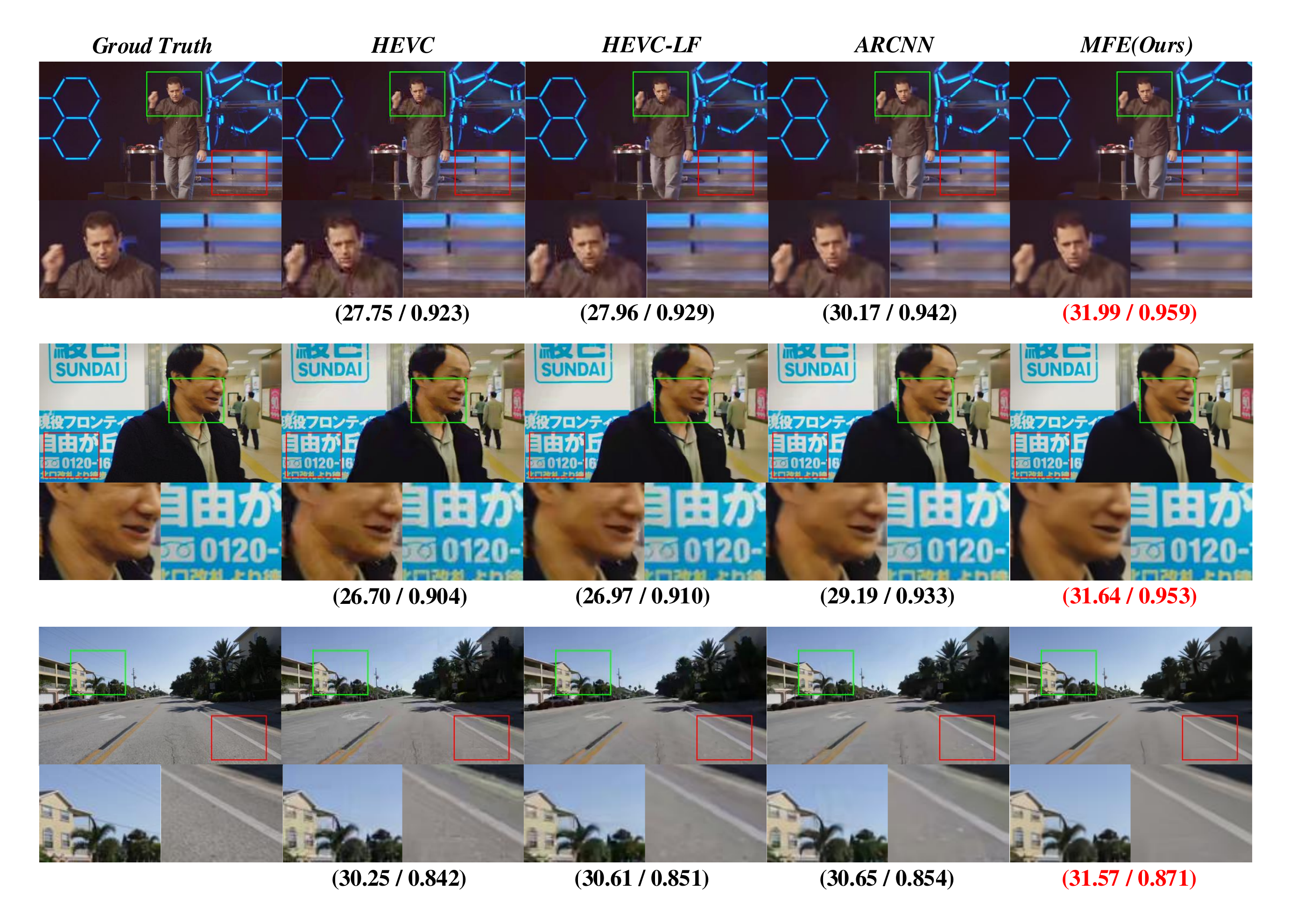}
\caption{Illustration of quantitative (PSNR/SSIM) and visual comparison of different methods for HEVC quality enhancement on the Vimeo dataset at QP=37.}
\label{sfig:compare}
\end{figure*}

%\subsection{Dataset \& experimental settings}
%\label{ssec:data&settings}

\subsection{Dataset}
\label{sssec:dataset}
To evaluate the performance of our proposed model, we chose the Vimeo-90K \cite{xue2017video} dataset which was built recently for different video processing tasks. A total of 4278 videos with 89800 independent clips at resolution of $448 \times 256$ were included. Among them, 64612 clips were used for training and 7824 clips for evaluation as the same as in~\cite{xue2017video}. We cropped the frames into the size of $256 \times 256$ randomly every time when the model read data during training to avoid the overflow of the memory.

\subsection{Model Training}
\label{ssec:training}
Our architecture is fully end-to-end trainable without requiring any component-wise pre-training. We train the model using the Adam optimizer \cite{kingma2014adam} with $\beta_1=0.9$, $\beta_2=0.999$ and $\epsilon=10^{-8}$. The learning rate is $10^{-4}$ initially, which is divided by 10 after each 20 epochs. The batch size is 4. Each batch is a set of 4 consecutive video frames, i.e., one I frame and three P frames.

We will not allow the loss $L_e$ back propagated to I frame to mimic the random access feature provided by the HEVC in this work. Thus, we use pre-trained SF QENet for each I frame. Additionally, we only use the nearest frame in decoding order as the reference for back-propagation to update model iteratively.  We firstly train our QENet on the dataset with QP 32 and utilize the training result as the pre-trained model to fine tune the model on the dataset with QP 37.

%Note that the loss $ L_{e} $ for restoration of the P frames mentioned in section 2.3 won't be back propagated to the I frame as explained in section 2.1. In other words, we use the pre-trained single-frame model to reconstruct the I frame without any parameter update during the training. In addition, we will only allow the losses back propagated to the previous frame instead of the two adjacent frames before and after to avoid occlusion introduced by merging two warped results. This is also in line with the requirement of low-delay coding applications with less computational complexity. Note that the losses will be back propagated to the nearest previous frame instead of the all previous frames to ensure the parameters of the model for each frame's reconstruction are updated once per training iteration, which prevents the weights of the losses from being different and results in effectiveness for enhancement of the whole sequences not just the first few frames. We firstly train our framework on the dataset with QP = 32 and utilize the training result as the pre-trained model to fine tune the model on the dataset with QP = 37.

\subsection{Performance Comparison}
\label{ssec:results}
To demonstrate the effectiveness of our proposed approach, we compare it with several typical artifact reduction methods regarding both single image and multi-frame sequences, i.e., ARCNN \cite{dong2014learning}, DnCNN \cite{zhang2017beyond}, DKFN \cite{lu2018deep} and default in-loop filters used in HEVC.

We have used well-known FFmpeg (\url{https://www.ffmpeg.org/}) with x265 (\url{https://x265.org}) enabled to
generate HEVC compliant bitstreams. Encoding parameters are kept as default, but with specific IPPP structure for low-delay scenario, and in-loop filters turned off (e.g., deblocking and SAO). This is referred to as the category ``HEVC'' that used for artifacts reduction efficiency comparison. Additionally, we provide another set of simulations with in-loop filters enabled and other parameters kept as the same, which is noted as category ``HEVC-LF''.  Such ``HEVC-LF'' versus ``HEVC'' could be used to tell the efficiency introduced by the default in-loop filters in HEVC.

%The results of HEVC-LF are generated by enabling deblocking filter and SAO in HEVC codec (libx265) using ffmpeg.
For ARCNN \cite{dong2014learning} and DnCNN \cite{zhang2017beyond}, we use the codes provided by the authors and train the models on the Vimeo training dataset. For DKFN  \cite{lu2018deep}, we directly cite their results provided in \cite{lu2018deep}. Vimeo evaluation videos are used to measure the efficiency of each methodology.

For fairness, we follow \cite{lu2018deep} and evaluate the $3^{rd}$ frame of each clip in the Vimeo dataset in terms of PSNR and Structural SIMilarity (SSIM)~\cite{wang2003multiscale}. Results are averaged for all evalution clips. The quantitative results are reported in Table~\ref{tab:results}. As seen, our proposed multi-frame approach, MF-QENet, presents the state-of-the-art performance, outperforming all existing models used in comparison, objectively. For QP 37, More than 1.5 dB PSNR gain is recorded for our MF-QENet over HEVC-LF; Even for DKFN, $>0.2$ dB PSNR gain is captured. Similar gains are kept for QP 32.
It is also observed that our single-frame approach, i.e., all the frames in the video are enhanced by just SF-QENet, exhibits close efficiency as the most recent multi-frame DKFN in~\cite{lu2018deep}.

Additionally, we have prepared visual comparison with HEVC-LF and ARCNN \cite{dong2015compression} to further evident the efficiency of our MF-QENet. As shown in Fig.~\ref{sfig:compare}, that artifacts caused by HEVC compression are clearly removed by our QENet, offering the superior subjective quality with smooth reconstruction, but noticeable blockiness are still kept even with other implementations.

\section{Concluding Remarks}
\label{sec:conclusion}
We propose an end-to-end trainable frame-recurrent quality enhancement framework as a standalone post-processing module for HEVC compliant low-delay compressed video applications. Three scale-wise convolutions are used in current frame to capture the multi-scale priors spatially and fused together to improve the spatial quality. In the meantime, temporal priors are introduced using predicted samples that are warped using extracted flows between consecutive decoded frames and previously enhanced frame, in a recurrent way.  As demonstrated in a public Vimeo dataset, our method has provided the state-of-the-art efficiency on artifacts reduction, against those popular and existing literatures, both objectively and subjectively.

The size of convolutions, as well as temporal priors in both forward and backward directions (e.g., B pictures) are our primary focuses for next step.

%\section{Acknowledgment}
%We would like thank all anonymous reviewers for their constructive comments to improve this manuscript.
%\vfill
\pagebreak
%-------------------------------------------------------------------------
\bibliographystyle{IEEEbib}
\bibliography{strings,refs}

\end{document}